\documentclass{statsoc}

\usepackage[a4paper]{geometry}
\usepackage{graphicx}
\usepackage[textwidth=8em,textsize=small]{todonotes}
\usepackage{amsmath, amssymb, mathtools}
\usepackage{natbib}
\usepackage{nameref,zref-xr}
\usepackage{url}
\usepackage{appendix}

\makeatletter

\title[Dominant Features in Finnish Forest Data]{Dominant-feature identification in data from Gaussian\\ processes applied to Finnish forest inventory records}
\author[Roman Flury {\it et al.}]{Roman Flury}
\address{Department of Mathematics, University of Zurich, Zurich, Switzerland.}
\email{roman.flury@math.uzh.ch}
\author{Tuomas Aakala}
\address{School of Forest Sciences, University of Eastern Finland, Joensuu, Finland.}
\author{Leena Ruha}
\address{Natural Resources Institute Finland (Luke), Oulu, Finland.}
\author{Timo Kuuluvainen}
\address{Department of Forest Sciences, University of Helsinki, Helsinki, Finland.}
\author[Flury R. et al.]{Reinhard Furrer}
\address{Department of Mathematics and Institute of Computational Science, University of Zurich, Zurich, Switzerland.}

\begin{document}

\section*{Abstract} 
In spatial data, location-dependent variation leads to connected structures known as features.
Variations occur at different spatial scales and possibly originate from distinct underlying processes.
Each of these scales is characterized by its own dominant features.
Here we introduce a statistical method for identifying these scales and their dominant features in data from Gaussian processes.
This identification involves credibly recognizing the dominant features by scale-space decomposition and assessing feature attributes by estimating covariance function parameters of the underlying processes and their associations to potential drivers.
We analyze Finnish forest inventory data from the 1920s using this dominant-feature identification method and identify the scales of variation in basal area estimates of most common Finnish trees, including Scots pine, Norway spruce, birch, and other native deciduous trees.
Comparing the resulting scale-dependent features and their attributes in these tree species, we identify the different effects of edaphic and anthropogenic drivers on the spatial distribution of their basal areas.
These data are analyzed for the first time in terms of their scale of variation, and the resulting scale-dependent maps and estimates are an essential contribution to the historical forest ecology of Fennoscandia.
Until now, this analysis was not possible with conventional methods.

\section{Introduction}
According to the first law of geography, everything is related to everything, but nearby things more so than distant things~\citep{Tobl:70}.
Nearby locations are therefore likely to have correlating values in spatial data and to form connected structures, which we term features.
Observed spatial data may be of different kinds across scales, resulting in distinct features at a given scale of variation.
Naturally, each scale is characterized by its own dominant features.
These scale-dependent features are usually the result of several underlying processes influenced by potentially different factors.
A key to understanding these processes is to recognize the individual scales and their dominant features.
Their identification makes it then possible to analyze more efficiently the underlying driving factors.
The entire process of detecting scales, recognizing their dominant features and evaluating their properties is known as \textit{dominant-feature identification}~\citep{Flur:Gerb:Schm:Furr:21}.

A variety of statistical models assume that spatially distributed data may originate from unknown underlying processes at multiple scales.
Prominent representatives are, among others, multi-scale Gaussian Markov random fields (GMRFs) \citep{Nych:Band:hamm:Lind:Sain:15} and multiresolution approximations of a spatial process covariance function for modeling and predictions that involve massively large data sets~\citep{Katz:17}.
The idea of multiresolution approximation was recently extended by~\cite{Paig:Fugl:Rieb:Wake:20} to 
involve Bayesian inference by integrated nested Laplace approximation (INLA)~\citep{Lind:Rue:Lind:11, Bakk:etal:18}.
\cite{Zamm:Roug:20} propose to model multi-scale data with non-stationary underlying processes using stacked processes.
These processes are thereby approximated with GMRFs to exploit the conditional dependence structure of the latent variables.
Most recently, \cite{Zhan:Katz:22} made use of the so-called Vecchia approximation for multi-scale processes, assuming that the scales are known a priori.

State-of-the-art, statistical scale-space multiresolution analysis~\citep{Holm:Pasa:Furr:Sain:11, Holm:Pasa:17} allows us to recognize the individual scales of variation in spatial data.
These approaches and their implementations have been optimized but remain limited to complete and regularly gridded data (pixel-based images).
\cite{Flur:Gerb:Schm:Furr:21} and \cite{Flur:Furr:19}~extended these ideas to spatial data assigned to a graph structure, including missing values, irregularly gridded and areal data using GMRFs.
However, the decomposition of geostatistical data without underlying grid or graph structure but originating from spatial processes has not yet been solved.
In Section~\ref{sec:featureidentification}, mathematical concepts and assumptions are outlined in detail to fill this gap in the scale-space literature.
Following the idea of~\cite{Pasa:Aaka:Holm2018} to determine the actual size of dominant features, we further develop the approach of~\cite{Flur:Gerb:Schm:Furr:21} for use with geostatistical data.
In addition, we propose modeling supplementary predictor variables to explain the variation that manifests in the same scale-dependent dominant features.

For forest structure and dynamics research, it is critical to identify the scales of variation, the effects of potential factors manifested in their spatial and scale-dependent features, and the changes in these scales over time.
In Section~\ref{sec:application}, we apply the outlined feature-detection method to Finnish forest inventory data from the 1920s.
These data has only been digitized recently, and compared, for example, to later forest inventories by~\cite{Hent:Nojd:Suva:Heik:Maki:20} and further analyzed by~\cite{Aaka:22}.
They bear potential insights regarding different spatial patterns of Finland's most common tree species prior to the onset of modern forestry.
Also, the scales of variation that produce different patterns are still unknown.
Differentiating between the edaphic and anthropogenic drivers is therefore of particular interest to understand the state and the underlying reasons at that time.
We use our method for dominant-feature identification to estimate their associations to scale-dependent features.
We provide the necessary context and background for this ecological application and discuss the results and draw statistically sound conclusions.
The results add to our understanding of how forests have developed in the past, and provide key evidence for understanding the development of forests prior to the onset of industrialized forestry.

\section{Feature identification}\label{sec:featureidentification}
In this section, we provide the details of the \textit{dominant-feature identification} method for geostatistical data.
We consider the situation where $\boldsymbol{y}$ is the observed data at $n \in \mathbb{N}^+$ conditionally independent and distinct locations $\boldsymbol{s} \in \mathcal{D} \subseteq \mathbb{R}^2$ and we assume that the observed data comprises of a realization $\boldsymbol{x} = (x_1, x_2, \ldots, x_n)^\top$ of a spatial process $\{ X(\boldsymbol{s}): \boldsymbol{s} \in \mathcal{D} \subseteq \mathbb{R}^2 \}$ and an orthogonal noise component.
We write $\boldsymbol{X}$ as the finite-dimensional representation of the process $X(\boldsymbol{s})$ observed at the $n$ locations.
Thus, we express the model as
\begin{equation}
    \boldsymbol{Y} = \boldsymbol{X} + \boldsymbol{\varepsilon}\label{eq:model}
\end{equation}
with $\boldsymbol{\varepsilon} \sim \mathcal{N} \left( \boldsymbol{0}, \boldsymbol{I}_n \sigma^2_{\boldsymbol{\varepsilon}} \right)$.
We assume that the spatial process is a zero-mean isotropic Gaussian process whose finite-dimensional covariance matrix $\boldsymbol{\Sigma}$ is defined through a covariance function $\text{COV}(X(\boldsymbol{s}_i), X(\boldsymbol{s}_j))=\text{cov}(||\boldsymbol{s}_i - \boldsymbol{s}_j||)$ for all  locations $\boldsymbol{s}_i,\boldsymbol{s}_j \in \mathcal{D}$ and additional covariance parameters $\boldsymbol{\theta}$ \citep{Cres:93}.
Throughout this paper, we use the Mat\'ern covariance function, unless specified otherwise, with the parametrization according to~\cite{Lind:Rue:Lind:11}.
Thereby, the covariance parameters $\boldsymbol{\theta}$ consist of $\rho > 0$, the spatial distance at which correlation is approximately $0.13$ (effective-range), of $\sigma > 0$, the marginal standard deviation (partial-sill) and of $\nu > 0$, the smoothness parameter, such that
\begin{equation}\label{eq:matern}
    \text{cov}(||\boldsymbol{s}_i - \boldsymbol{s}_j||) = \frac{\sigma^2}{2^{\nu - 1} \Gamma(\nu)} (\sqrt{8\nu}/\rho ||\boldsymbol{s}_i - \boldsymbol{s}_j||)^{\nu} \text{K}_{\nu} (\sqrt{8\nu}/\rho ||\boldsymbol{s}_i - \boldsymbol{s}_j||).
\end{equation}
Here, $\Gamma$ is the Gamma function and $\text{K}_\nu$ is the modified Bessel function of the second kind of order $\nu$.

We use the likelihood function of the Gaussian process to estimate covariance function parameters.
However, other estimation procedures, such as low-rank approximations, can be used instead.
The following derivations are analogous to (non-stationary) covariance models other than that of Mat\'ern.
Also, these derivations are not restricted to data following Gaussian processes.
Other processes are applicable, e.g., data following a Poisson process~\citep{Agar:Gelf:Citr:02} or using a suitable link function~\citep{Mill:Glen:Seat:20}.
Moreover, there are no restrictions on the surface of the domain $\mathcal{D} \subset \mathbb{R}^2$ and its respective coordinate reference systems as long as a well-defined distance measure is used.

\subsection{Field reconstruction}\label{sec:fieldreconstruction}
After Model~\eqref{eq:model}, we assume the observed data is a realization of a composition of $\boldsymbol{X}$ and some observational or measurement noise $\boldsymbol{\varepsilon}$.
In order to separate the noise component from the observed field $\boldsymbol{y}$, we sample from $\boldsymbol{X}$ and approximate $\boldsymbol{x}$ with the mean of sufficient sampling draws.
As sampling approach, we propose so-called conditional sampling. 
Thereby we are using that $\boldsymbol{X}$ and $\boldsymbol{\varepsilon}$ follow both a zero-mean Gaussian distribution, which implies that $\boldsymbol{Y} \sim \mathcal{N}_n\left(\boldsymbol{0}, \boldsymbol{\Sigma} + \sigma^2_{\boldsymbol{\varepsilon}}\boldsymbol{I}_n \right)$.
Furthermore, it holds that
\begin{equation}
\begin{pmatrix}\boldsymbol{X}\\
              \boldsymbol{Y} \end{pmatrix} \sim \mathcal{N}_{2n}
              \left( \boldsymbol{0}, \begin{pmatrix*}[l]
                        \boldsymbol{\Sigma} \quad & \boldsymbol{\Sigma}\\
                        \boldsymbol{\Sigma} \quad & \boldsymbol{\Sigma}+\sigma^2_{\boldsymbol{\varepsilon}}\boldsymbol{I}_n 
              \end{pmatrix*} \right)\label{eq:2ndistr}.
\end{equation}
Moreover, as $\boldsymbol{\varepsilon}$ is assumed to be independent random noise, $\boldsymbol{X}$ and $\boldsymbol{\varepsilon}$ are independent and the cross-covariance submatrices are equivalent to $\boldsymbol{\Sigma}$.
The conditional distribution $\boldsymbol{X}|\boldsymbol{y} \sim \mathcal{N}_n \left( \boldsymbol{\Sigma}(\boldsymbol{\Sigma} + \sigma^2_{\boldsymbol{\varepsilon}}\boldsymbol{I}_n)^{-1}\boldsymbol{y}, \boldsymbol{\Sigma} - \boldsymbol{\Sigma}(\boldsymbol{\Sigma} + \sigma^2_{\boldsymbol{\varepsilon}}\boldsymbol{I}_n)^{-1}\boldsymbol{\Sigma}  \right)$ then follows again a multivariate Gaussian distribution.
Using a maximum likelihood (ML) approach, we estimate the parameters of the covariance function defined in Equation~\eqref{eq:matern} and the variances of the noise $\sigma^2_{\boldsymbol{\varepsilon}}$ with the observed data $\boldsymbol{y}$.

For efficient sampling from this conditional distribution, we can apply the following steps:
\begin{enumerate}
  \item Sample realizations of $\boldsymbol{Z} = (\boldsymbol{X}, \boldsymbol{Y})^\top$ using the density function and covariance matrix according to Equation~\eqref{eq:2ndistr}.
  \item Transform the realizations from $\boldsymbol{Z}$ to $\boldsymbol{X}|\boldsymbol{Y}$ with $\boldsymbol{X} + \boldsymbol{\Sigma}(\boldsymbol{\Sigma} + \sigma^2_{\boldsymbol{\varepsilon}}\boldsymbol{I}_n)^{-1}(\boldsymbol{y} - \boldsymbol{Y})$.
\end{enumerate}
As a result, only one additional linear system based on $(\boldsymbol{\Sigma} + \sigma^2_{\boldsymbol{\varepsilon}}\boldsymbol{I}_n)$ needs to be solved.
Approximate conditional sampling approaches can be applied if necessary, for example as described in~\cite{Bail:Band:Nych:21} or, for observations on the sphere, by~\cite{Emer:Furr:Porc:19}.
Alternatively, if prior knowledge is available about the support of the sample space, a Bayesian hierarchical model~\citep{Gelf:12} can be used instead of conditional sampling.
In this case, $\text{E}[\boldsymbol{X}|\boldsymbol{y}]$ denotes the posterior sample mean.

\subsection{Decomposition}
In classical scale-space analysis, a roughness penalty smoother is commonly used to smooth the data on multiple scales, so that a decomposition can be found \citep{Holm:Pasa:Furr:Sain:11}.
In general, a roughness penalty smoother  is defined as $\boldsymbol{S}_{\lambda} = (\boldsymbol{I}_n + \lambda \boldsymbol{Q})^{-1}$, where $\lambda$ is the smoothing scale and $\boldsymbol{Q}$ is a spatial weight matrix.
As we introduced a spatial process in Model~\eqref{eq:model}, we express a smoother in terms of the correlation matrix $\boldsymbol{R}$, such that the entries of $\sigma^2\boldsymbol{R}$ are given by Equation~\eqref{eq:matern}, with
\begin{equation}
    \boldsymbol{\widetilde{S}}_{\lambda} = \boldsymbol{R} (\boldsymbol{R} + \lambda\boldsymbol{I}_n)^{-1}.\label{eq:slambdatilde}
\end{equation}
When $\boldsymbol{Q}$ is the precision matrix $(\boldsymbol{I}_n + \lambda \boldsymbol{Q})^{-1}=1/\lambda \boldsymbol{Q}^{-1}(1/\lambda \boldsymbol{Q}^{-1} + \boldsymbol{I}_n)^{-1} = \sigma^2\boldsymbol{R} (\sigma^2\boldsymbol{R} + \lambda\boldsymbol{I}_n)^{-1}$, this smoother is conceptually equivalent to the roughness penalty smoother.
Analogous to the classical definition of the penalty smoother, it holds that $\lim_{\lambda \to 0} \boldsymbol{\widetilde{S}}_{\lambda}\boldsymbol{x} = 
\boldsymbol{x}$,  and $\lim_{\lambda \to \infty} \boldsymbol{\widetilde{S}}_{\lambda}\boldsymbol{x} =
 \boldsymbol{0}$.
Thereby $1/\lambda$ can be compared to the \textit{signal to noise} ratio.

The Mat\'ern smoothing correlation matrix $\boldsymbol{R}$ in the definition of $\boldsymbol{\widetilde{S}}_{\lambda}$ is parametrized by individual parameters that are tuning parameters of this smoother.
We propose to choose the range sufficiently large based on prior knowledge of the data, e.g., topographical attributes.
The smoothness parameter needs to be chosen large enough such that it represents the smoothness of $\boldsymbol{x}$ adequately.
To show how different range and smoothness configurations influence this decomposition, we provide a simulation study in Supplementary Material~\ref{sec:scaleselection}.

To decompose $\boldsymbol{x}$, we consider a sequence of smoothing scales $0 = \lambda_1 < \lambda_2 < \ldots < \lambda_{L} = \infty$ and apply it to $\boldsymbol{\widetilde{S}}_{\lambda}\boldsymbol{x}$. 
Thereby, $\boldsymbol{\widetilde{S}}_{0}\boldsymbol{x} = \boldsymbol{x}$ and $\boldsymbol{\widetilde{S}}_{\infty}\boldsymbol{x} = 0$ by assumption of the overall mean of $\boldsymbol{x}$.
Then $\boldsymbol{x}$ can be represented as: $\boldsymbol{x} = \boldsymbol{\widetilde{S}}_{0}\boldsymbol{x} - \boldsymbol{\widetilde{S}}_{\lambda_2}\boldsymbol{x} + \boldsymbol{\widetilde{S}}_{\lambda_2}\boldsymbol{x} - + \ldots - \boldsymbol{\widetilde{S}}_{L-1}\boldsymbol{x} + \boldsymbol{\widetilde{S}}_{L-1}\boldsymbol{x}$.
Scale-dependent details are formalized as differences between consecutive smooths $\boldsymbol{z}_{\ell} = (\boldsymbol{\widetilde{S}}_{\lambda_{\ell}} - \boldsymbol{\widetilde{S}}_{\lambda_{\ell+1}} )\boldsymbol{x}$ 
for $\ell = 1, \ldots, L-1$.
That is, we express $\boldsymbol{x}$ as a sum of these details $\boldsymbol{x} = \sum_{\ell = 1}^{L-1} \boldsymbol{z}_{\ell}$.
In practice this decomposition is calculated for each sample draw from Section~\ref{sec:fieldreconstruction} and the details are summarized by their sample mean $\text{E}[\boldsymbol{z}_{\ell}|\boldsymbol{y}]$.
In the following we simplify notation and denote the summarized details by $\boldsymbol{z}_{\ell}$.

\subsection{Scale selection}
In order to select sensible smoothing scales, we follow~\cite{Pasa:Laun:Holm:13}, who introduced the idea of scale derivatives that can be adapted for the spatial smoother $\boldsymbol{\widetilde{S}}_{\lambda}\boldsymbol{x}$, analogously defined by $\widetilde{\boldsymbol{D}}_{\lambda}\boldsymbol{x} = \frac{\partial \boldsymbol{\widetilde{S}}_{\lambda}}{\partial \log\lambda}\boldsymbol{x}$.
As the degree of smoothing increases, the difference between successive values for $\lambda$ has to become larger and more significant to have a noticeable effect on the smoothing, which is why the scale derivative $\lambda$ is on a logarithmic scale.
The scale derivative can be expressed in terms of the smoothing correlation matrix $\boldsymbol{R}$ as
\begin{equation}
    \widetilde{\boldsymbol{D}}_{\lambda} \boldsymbol{x} = -\lambda \boldsymbol{R} (\boldsymbol{R} + \lambda\mathbf{I}_n)^{-1} (\boldsymbol{R} + \lambda\mathbf{I}_n)^{-1}\boldsymbol{x}.\label{eq:scalederivativetilde}
\end{equation}
This scale derivative shows the change of the field in dependence of the smoothing scale.
We choose $\lambda$ corresponding local minima of this scale derivative with respect to the Euclidean or maximum norm.
In this manner, the difference of the corresponding smoother captures all scale-dependent dominant features related to the maximum between respective minima.
\cite{Flur:Gerb:Schm:Furr:21} showed that the maximum norm is more sensitive and can detect local extremes that may arise from anisotropic or non-stationary spatial processes.
Data may not be clearly separable or local variations in large data sets might be smoothed out.
The sensitivity of the maximum norm is therefore a crucial advantage.

\subsection{Credibility analysis}
To credibly recognize the dominant scale-dependent features, we calculate probabilities based on the individual samples from the field reconstruction (Section~\ref{sec:fieldreconstruction}) and the derived smoothing scales.
Here, we use so-called pointwise (PW) credibility maps to recognize credible dominant features, i.e., the areas where the component is credibly positive or negative.
For the PW map of the $\ell$-th detail $\boldsymbol{z}_{\ell}$, considered as a realization of the random vector $\boldsymbol{Z}_{\ell}$, each location $\boldsymbol{s}_i \in \mathcal{D}$ is allocated either to subsets $I^\text{high}=\{ i: \text{P}(\boldsymbol{Z}_{\ell, \boldsymbol{s}_i} > 0\mid \boldsymbol{y}) \ge \alpha \}$ or $I^\text{low}=\{ i: \text{P}(\boldsymbol{Z}_{\ell, \boldsymbol{s}_i} < 0\mid \boldsymbol{y}) \ge \alpha \}$, in which the detail $\boldsymbol{z}_{\ell,s}$ locationwise differs credibly from zero with respect to the credibility level $\alpha$, typically $\alpha = 95\%$.
If  a $\boldsymbol{z}_{\ell, \boldsymbol{s}_i}$ is neither allocated to $I^\text{high}$ nor $I^\text{low}$, it is allocated to $I^\text{null}=\{1,\dots,n\}\setminus \left(I^\text{high} \cup I^\text{low}\right)$.
In PW maps, probabilities are calculated independently for each location. Other credibility maps, such as \textit{highest pointwise probabilities} or \textit{simultaneous credible intervals}, increase the cohesiveness of credible locations~\citep{Era:Holm:05, Holm:Pasa:Furr:Sain:11}.
Furthermore, \cite{Boli:Lind:15} proposes to estimate \textit{excursions} and \textit{contour uncertainty regions} for latent Gaussian models based on a parametric family for the excursion sets combined with posterior samples.

\subsection{Feature attributes}
In order to complete the dominant-feature identification method for geostatistical data, we assess the characteristics of each detail.
Therefore we estimate spatial properties and, as a new aspect, also scale-dependent linear fixed effects of additional predictor variables.
We denote the whole set of characterizing parameters and linear effects as feature attributes.

We can model the details $\boldsymbol{z}_{\ell}$ for $\ell=1,\, \ldots, \, L-1$ either as pure spatial components or conditional on linear effects of $k$ predictor variables.
In the former case, we use Gaussian processes $V_{\ell}(\boldsymbol{s})$, so that
\begin{equation}
    \boldsymbol{z}_{\ell} \text{ a finite dimensional realization of } V_{\ell}(\boldsymbol{s}), \quad  \boldsymbol{s} \in \mathcal{D}.\label{eq:purespatialmodel}
\end{equation}
In the latter case, we have
\begin{equation}
    \boldsymbol{z}_{\ell} \text{ a finite dimensional realization of } \boldsymbol{w}_{\ell}(\boldsymbol{s})^\top\boldsymbol{\beta}_{\ell} + V_{\ell}(\boldsymbol{s}), \quad  \boldsymbol{s} \in \mathcal{D},\label{eq:linearspatialmodel}
\end{equation}
where $\boldsymbol{w}_{\ell}(\boldsymbol{s})$ is the vector of scale-dependent predictors at locations $\boldsymbol{s}$.
The design matrix $\boldsymbol{W}\kern-3pt_{\ell} \in \mathbb{R}^{n \times k}$ contains the rows $\boldsymbol{w}_{\ell}(\boldsymbol{s})^\top$ for $i = 1, \ldots, n$, and $\boldsymbol{\beta}_{\ell} \in \mathbb{R}^k$ the linear coefficients.
We propose to construct the scale-dependent design matrix by decomposing the predictor variables with the derived smoothing scales for $\boldsymbol{x}$.
In particular for predictors containing robust small- and large-scale features, this potentially leads to essential insights that cannot be obtained using un-decomposed predictors~\citep{Pasa:Holm:17}.

From a marginal point of view, $\boldsymbol{z}_{\ell}$ is a realization of a zero-mean Gaussian process $V_{\ell}(\boldsymbol{s})$, whose finite dimensional covariance matrix is $\boldsymbol{\Sigma}_{\ell}$.
We can decompose $\boldsymbol{\Sigma}_{\ell}=\boldsymbol{\Sigma}_{\ell, \text{fixed}} + \boldsymbol{\Sigma}_{\ell, \text{random}}$, according to components from selected covariates with design matrix $\boldsymbol{W}\kern-3pt_{\ell}$ of linear predictor variables and $\boldsymbol{\beta}_{\ell}$ linear coefficients; i.e., $\boldsymbol{\Sigma}_{\ell, \text{fixed}} = \text{VAR}(\boldsymbol{W}\kern-3pt_{\ell}\boldsymbol{\beta}_{\ell})$, the variance of the fixed effects.
For estimation, we place ourselves in a conditional framework and assume for the detail $\boldsymbol{z}_{\ell}$ the model $\mathcal{N} (\boldsymbol{W}\kern-3pt_{\ell}\boldsymbol{\beta}_{\ell}, \boldsymbol{\Sigma}_{\ell, \text{random}})$.
When thereby using an ML approach to estimate the set of feature attributes and when data size is moderate, the Hessian matrix can be calculated to derive Wald-confidence intervals for each estimate.

\subsection{Overfitting}
Overfitting is a common problem with statistical models and it can arise when identifying dominant features in spatial data.
Choosing too many scales leads to details containing remarkably similar scale-dependent features.
Moreover, while estimating feature attributes, overfitting may arises when the model describes the random error in a detail rather than the associations to explanatory variables.
We propose spatial cross-validation to address these issues.
We divide the locations $\boldsymbol{s} \in \mathcal{D}$ into different subareas of approximately equal shape and size; for example, into rectangular-shaped blocks.
Furthermore, we assign these blocks to $k$ different training sets, each leaving approximately $1/k$-th of the data out.
We identify dominant features for each training separately and analyze whether the scale derivative and the selected scales are similar compared to the scales based on the entire data.
Local extrema may affect the analysis if the training set suggests remarkably different scales.
This could also indicate that the assumption of isotropy is inappropriate.

Overfitting can be quantified by comparing predictions from Models~\eqref{eq:purespatialmodel} or~\eqref{eq:linearspatialmodel} using, on the one hand, the estimates based on the training sets and, on the other hand, the respective test-set locations from the sum of all details~$\boldsymbol{z}_{\ell}$.
Classical measures and scores, such as the root-mean-squared error (RMSE) and the continuous ranked probability score (CRPS)~\citep{Gnei:Bala:Raft:07}, can be calculated.

Additional, but herein not used approaches to tackle overfitting are to use the generalized likelihood ratio hypothesis test, i.e.,~to test whether the partial sill $\sigma^2_{\ell}$ for each detail process is significantly different from zero, or to use a regularized likelihood approach, broadly in the spirit of \cite{Damb:Sigr:Furr:21}.

\subsection{Summary}
In Section~\ref{sec:application} we apply the dominant-feature identification method to scientifically relevant data. Before doing so, we summarize here the steps:
\begin{enumerate}
  \item Estimate the covariance function of $\boldsymbol{Y}$ to draw samples of $\boldsymbol{X}|\boldsymbol{y}$, separating noise from observations $\boldsymbol{y}$.
  \item Determine smoothing scales used for decomposing $\boldsymbol{x}$ with scale derivatives, based on the sample mean of $\boldsymbol{X}|\boldsymbol{y}$.
  \item Calculate details based on individual sample draws of $\boldsymbol{X}|\boldsymbol{y}$ and summarize by their sample mean, simultaneously deriving PW probability maps.
  \item Assess feature attributes for each detail $\boldsymbol{z}_{\ell}$ by estimating linear and spatial effects by ML.
  \item Estimate overfitting by repeating the previous steps for $k$ subsets.
\end{enumerate}

\section{Identifying dominant features in Finnish forest inventory records}\label{sec:application}
Natural processes and human activities change forest structures and species compositions.
These changes have generally experienced an increase over the past centuries, with increasing human population, variety of forest uses, and later by the rise of the forest industries.
While a variety of data sources, such as remote sensing and forest inventory data, give us a good understanding of how forests are changing in managed and protected areas in intensively managed regions such as Fennoscandia~\citep{Tomp:etal:08}, the interplay between natural forest development and the influence of human actions prior to modern forestry is much less well known.
In particular, the spatial patterns across large regions and the scales at which they vary are unknown.
Identifying these patterns and scales would further aid in understanding the drivers behind these patterns.
Analyzing such scientific questions is possible for the Finnish forest by considering available nationwide forest inventories completed by field crews in the last century.

The first systematic nationwide forest inventory (NFI1) in Finland was conducted between 1921 and 1924~\citep{Ilve:27}.
The aim was to provide a reliable statistical description of the forest and tree stocks.
Following the digitalization of these data, recent studies have complemented the picture that emerged from the original data, with a focus on the development of the amount of large and/or old trees and on tree-size distributions in different  regions~\citep{Hent:Nojd:Suva:Heik:Maki:19, Hent:Nojd:Suva:Heik:Maki:20}.
\cite{Aaka:22}~recently developed interpolated maps based on a subjective selection of the degree of smoothing to demonstrate larger-scale variability in several forest characteristics without relying on regions determined a priori.
However, whether the regional division or the subjectively chosen smoothers correspond to actual scales of variation in the data remained unknown.
By applying the outlined identification of dominant features to these NFI1 data, we can identify these scales of variation and compare the scale-dependent features among different tree species, taking into account the influence of edaphic and anthropogenic variables.

\subsection{Ecolocigal context}
Finland is situated roughly between latitudes 60 and 70 North and longitudes 21 and 33 East.
Climatic conditions correlate strongly with changes in latitude.
For example, the length of the growing season reflects this correlation, which in the forested part of the country ranges from 180 days in Helsinki in southern Finland to only 120 days in Sodankylä in the north.
The majority of the Finnish landscape is forested, with boreal forests currently covering more than 70\% of the country.
The most common tree species in Finland are Scots pine (\textit{Pinus sylvestris}), Norway spruce (\textit{Picea abies}), and birch (\textit{Betula pendula} and \textit{B.~pubescens}).
Other native deciduous trees such as aspen and alder complete the tree stock of Finland.
A common attribute assessed for tree stocks is the basal area (BA), which is a structural stand characteristic typically used to describe forest stand density and a proxy for timber volume or growth.
BA is measured as the amount of area occupied by tree stems, per unit area, typically expressed as $\text{m}^2/\text{ha}$.
BA is influenced by site productivity (soil, topography, climate), forest age, species compositions and disturbance history (either natural disturbances or logging).
When developing naturally, the species dominance is mainly influenced by site productivity and the occurrence of fires.
Forests on productive sites (both well- and poorly-drained) typically develop into spruce-dominated stands after an initial dominance of deciduous trees.
Conversely, on low-productive and dry sites, pine tends to dominate.
On intermediate sites, absence of fires increases dominance of spruce, which is a fire-intolerant species.
Conversely, fire occurrence tends to favor pines on these sites.

Regarding the drivers of variation in BA, we therefore hypothesize that at small scales, differences in BAs of different species are driven by differences in edaphic conditions (site-type), but at larger scales, after controlling for climatic influence, this variation reflects geographical variation in how people used the forests.
First, we expect the commonness of slash-\&-burn agriculture to lead to a decline in spruce and an increase in deciduous trees.
This is because the more productive spruce-dominated sites are primary areas for slash-\&-burn agriculture. 
The areas typically regenerate with deciduous trees following the abandonment of cultivated areas.
Second, forest grazing tends to favor other deciduous trees, including species such as alder, which is unpalatable to cattle.
Third, population density reflects the pressure for household consumption of wood, especially for fuel, but also for construction and other material uses.
With that in mind, we expect especially pine to show a negative relationship with human influence.

\subsection{NFI1 data}
Testing the hypothesis outlined above, we consider the NFI1 data and other historical data sources described in the following.
For this foremost forest inventory, field crews took tree samples on inventory lines from southwest to northeast across Finland, with a 26~km distance between most lines (for details, see~\citealp{Ilve:27, Hent:Nojd:Suva:Heik:Maki:19}).
The cumulative length of these lines is 13'348~km, without considering areas covered by water.
At these inventory lines, sample plots of size $10\,\text{m} \times 50\,\text{m}$ were considered to assess land use and site characteristics.
In the original outline, the distance between plots was 2~km; however, we shifted plots so that each plot contained an entire forest stand.

\begin{figure}[ht]
  \centerline{\includegraphics[width=\textwidth]{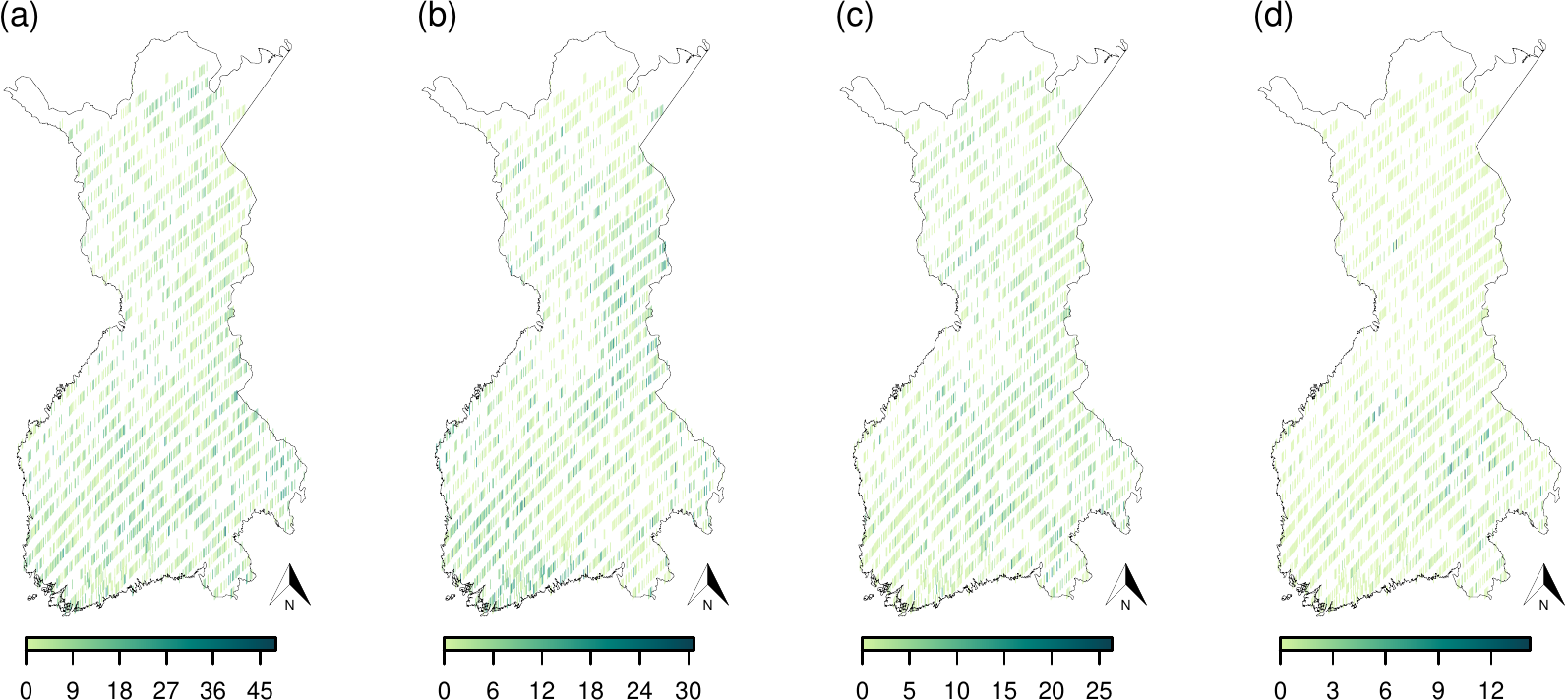}}
  \caption{\small BA based on NFI1:
           for (a) pine; (b) spruce; (c) birch; (d) other deciduous trees.}
  \label{fig:fl_data}
\end{figure}
The NFI1 field crews measured the diameter at breast height (\/\textit{dbh}), 1.3~m above ground, of each tree for all $10\,\text{m} \times 50\,\text{m}$ sampling plots.
Trees were classified into 2~cm classes so that trees larger than 4~cm were consistently recorded in all plots (4--6~cm, 6--8~cm, and so on).
We computed the plot-level BA, based on the \textit{dbh} measures, by transforming the \textit{dbh} to the area of a circle; that is, $\text{BA} = \pi \cdot (dbh/2)^2$.
In the NFI1 data, the \textit{dbh} measures are separately available for pine, spruce, birch, alder, aspen, and other broadleaf trees, where the last three are summarized as other deciduous trees; cf.~panels (a) to (d) in Figure~\ref{fig:fl_data}.
In total, the data set contained 3'065 BA estimates on forests on mineral soil.
A visual inspection of the BAs in Figure~\ref{fig:fl_data} implies high and uniform occurrence of pine throughout Finland compared to all other tree species.
The second-highest values are evident in the BA of spruce; however, these appear to be less evenly distributed, and in the southeastern part of the country, there is a very low occurrence of spruce.
Based on the BAs of birch, a subordinate occurrence of birch seems to be a fixed component of Finnish forests, with some exceptions where birch are more dominant in forest stands.
In particular, birch may dominate forests in the North, close to treeline.
Other deciduous trees were abundant only in the southern half of Finland.

\begin{figure}[ht]
  \centerline{\includegraphics[width=\textwidth]{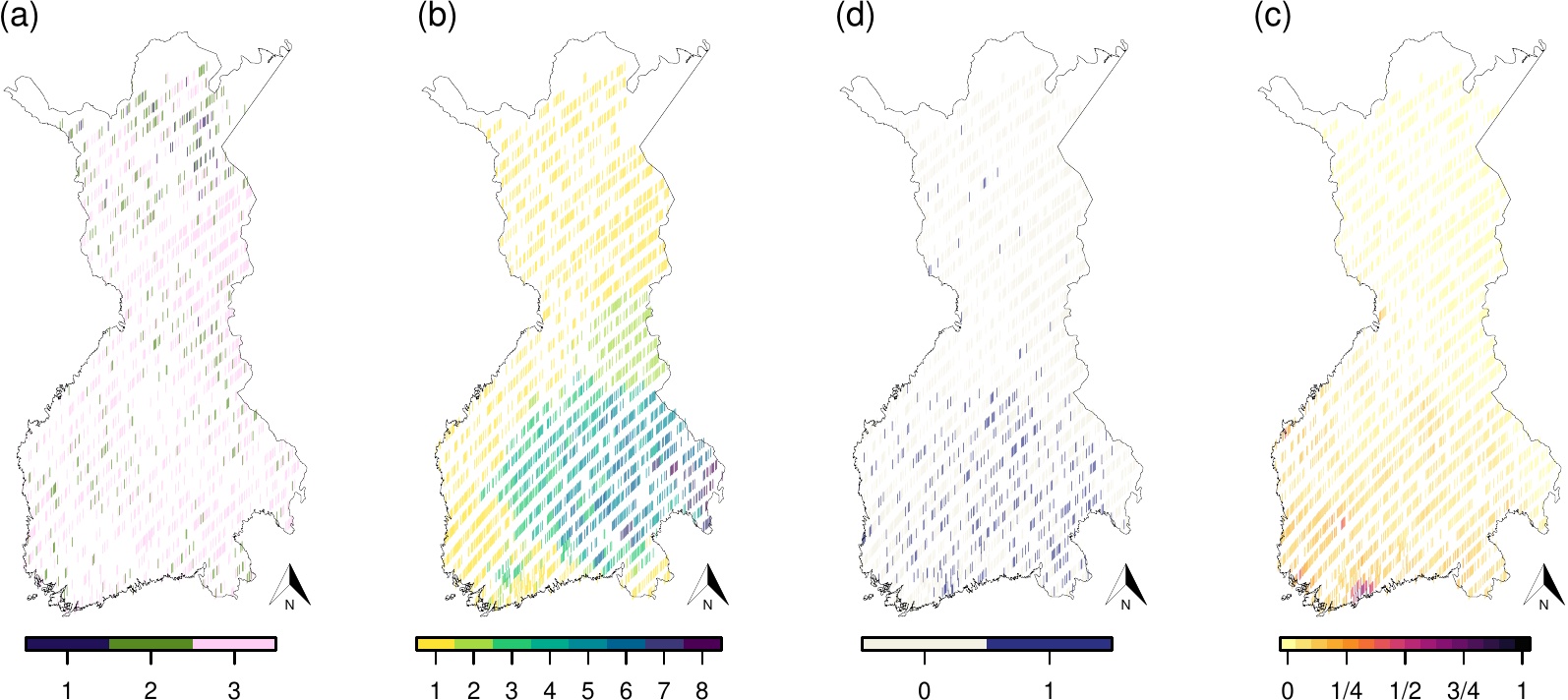}}
  \caption{\small Edaphic and anthropogenic variables: 
           (a) site-type, where the categories \textit{xeric} (1), \textit{sub-xeric} (2), \textit{mesic} (3), correspond to the assigned levels; (b) slash-\&-burn agriculture intensity levels; (c) no-grazing or grazing; (d) population density (min-max normalized).}
  \label{fig:fl_data2}
\end{figure}
During NFI1 sampling, site type was determined for each plot based on a visual assessment.
The corresponding productivity classes cover this edaphic influence on the forest stands.
We grouped the detailed classification into three broader classes \textit{xeric}, \textit{sub-xeric} and \textit{mesic}, cf.~panel (a) of Figure~\ref{fig:fl_data2} and~\cite{Ilve:27}. 
Originally, the Finnish equivalent classes from barren to herb-rich were assessed: \textit{karukko}, \textit{kuiva}, \textit{kuivahko}, \textit{tuore} and \textit{lehtomainen}.
However, \textit{karukko} and \textit{kuiva} as well as \textit{tuore} and \textit{lehtomainen} were combined with xeric and mesic respectively, given that there were only few observations.
The most dry-barren and dry sites were found in the very northern part of Finland.

We considered the following three anthropogenic variables to account for human influence on the forests.
First, the prevalence of slash-\&-burn agriculture, a method of cultivation in which the trees in a forest area are burned and cleared for several years of cultivation, followed by a few decades of rest to let the forest recover.
The intensity of this practice in 1913 in Finland is shown in panel (b) of Figure~\ref{fig:fl_data2}; the data origins from~\cite{Heik:15}.
The more intensively this practice was used over these years, the higher the level. 
In the past century, this practice was used most commonly in the southeast of Finland, where the spruce occurrence is very low, and other deciduous trees are most abundant.
Second, the binary variable grazed describes whether the respective plot was used for forest grazing; cf.~panel (c) in Figure~\ref{fig:fl_data2}.
Often, cattle were let into forests to graze, which were left to recover from slash-\&-burn agriculture cultivation independent of the intensity of this practice.
In addition we consider the population density in 1925 of Finland; cf.~panel (c) in Figure~\ref{fig:fl_data2}.
The continuous values for each plot location are based on an inverse-distance weighted interpolation of population density~\citep{Aaka:22}, from the digitalization of a settlement map from~\cite{Witt:28}. 
The figure shows that Finland is more populated in the south, and the larger cities are concentrated in the southwest.

\subsection{Dominant-feature identification steps}
In this section, we describe the individual steps of the dominant-feature identification procedure for geostatistical data presented in Section~\ref{sec:featureidentification}, applied to the described BAs of the NFI1 data.
The computational steps are implemented in the statistical software \textsc{R}~\citep{R}, and are openly available in the repository associated with this manuscript; cf. Section~\ref{appen:code} of the Supplementary Material for details.
All steps are equivalently applied to BA of pine, spruce, birch, and other deciduous trees (other).
These four datasets include 3'065 data points each, which determines the dimensions of the respective covariance matrix structures.
The sizes of the corresponding computational objects are moderate.
We apply spatial cross-validation to ensure that the scales and final estimates are not a result of overfitting.
Therefore we blockwise (of size $26$~km$ \times 26$~km) divide the area of interest into five subsets of approximately 80\% of the data and repeat the described steps for each subset.

First, we preprocess the BA data by removing the natural linear trend in north--south direction as the growth rates of the trees depend on their latitude coordinate, with a stepwise decline along these.
Omitting to detrend would imply an additional scale and detail in the decomposition.
Hence to improve separability and considering that the trend based on the growth rates is not the focus of the outlined hypothesis, we remove this trend.
Subsequently, we continue with the standardized residuals of these detrended BAs, which are assumed to follow a zero-mean Gaussian distribution.
Furthermore, we assume the data is a realization of an isotropic spatial process, making the dominant-feature identification method applicable.
Next, we make use of the conditional sampling approach to remove observational and measurement noise $\boldsymbol{\varepsilon}$ from the observed data $\boldsymbol{y}$ according to Model~\eqref{eq:model}.
Therefore, we draw 1'000 random samples and approximate $\boldsymbol{x}$ with the average of these samples.

\begin{figure}[ht]
  \centerline{\includegraphics[width=\textwidth]{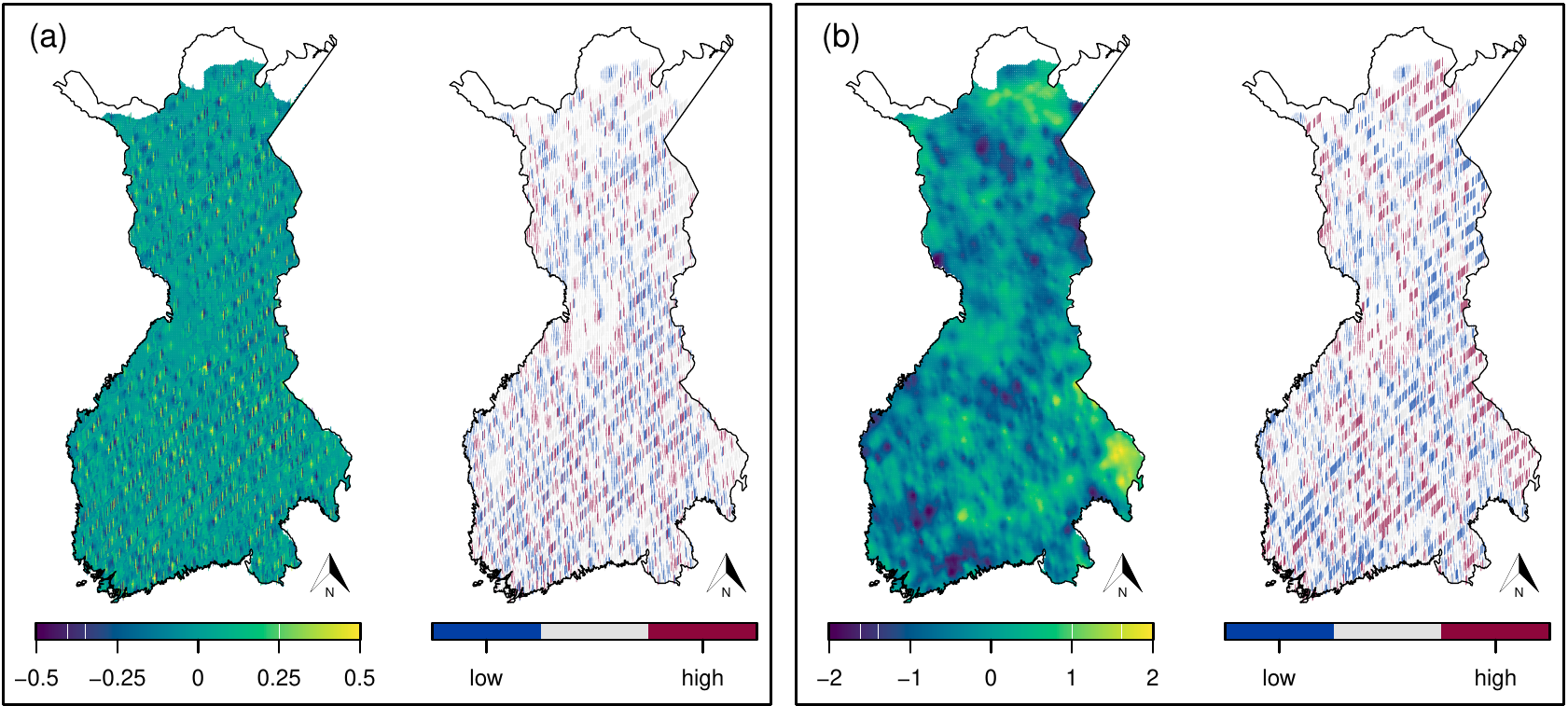}}
  \caption{\small Interpolated scale-space decomposition of pine BA: (a) pine $\boldsymbol{z}_1$ and PW$_1$; (b) pine $\boldsymbol{z}_2$ and PW$_2$.}
  \label{fig:fl_pine}
\end{figure}
To select the scales, we are using the scale derivative defined in Equation~\eqref{eq:scalederivativetilde}, thereby calculating the smoothing correlation matrix according to a Mat\'ern covariance function (Equation~\eqref{eq:matern}).
We set the parameters of this function such that the effective range $\rho$ corresponds to 26~km, ensuring that multiple inventory lines are considered to model the smoothing correlation matrix and $\nu = 0.5$ to reflect the roughness in the data.
The smoothing correlation matrix is computed based on great-circle distances between all plot locations.
Henceforth we can select the scales according to the minimas of the maximum norm of the scale derivative.
In all four BA data sets, we can identify one minimum leading to two details.
The scales are (standard deviation of cross-validation results in parentheses) $\lambda_{\text{pine}} = 1,\, (1.16)$, $\lambda_{\text{spruce}} = 0.47,\, (0.45)$, $\lambda_{\text{birch}} = 1.73,\, (0.44)$, and $\lambda_{\text{other}} = 0.27,\, (0.08)$.

In the next step, we calculate the decomposition of the BAs, based on the selected smoothing scales, the smoothing correlation matrix and the individual sampling draws.
We obtain for each BA two respective details summarized with their conditional sample means, i.e.,~pine/spruce/birch/other $\boldsymbol{z}_1$ and~$\boldsymbol{z}_2$.
We use PW probability maps based on the individual sampling draws from the sampling step to visualize dominant features more clearly.
Figure~\ref{fig:fl_pine} shows both details and PW maps for pine BA and respective figures in Section~\ref{app:flplots} of the Supplementary Material show the decomposition for BAs of spruce, birch, and other trees.
The respective detail and PW maps are interpolated to the whole of Finland, where a weighted $k$-nearest neighbors approach is used to complete the predictor variables~\cite{Hech:Schl:04}.

We assess dominant scale-dependent feature attributes by simultaneously estimating the spatial correlation and the effects of the described edaphic and anthropogenic drivers.
Therefore, we use for the respective two details Model~\eqref{eq:linearspatialmodel}, where the scale-dependent design matrices $\boldsymbol{W}\kern-3pt_1$ and $\boldsymbol{W}\kern-3pt_2$ are constructed with the decomposed driver variables, $\boldsymbol{\beta}_1$ and $\boldsymbol{\beta}_2$ are the vectors of corresponding scale-dependent linear coefficients, and $V_{1}(s)$, $V_{2}(s)$ are the underlying zero-mean isotropic Gaussian processes, characterized with a Mat\'ern covariance function.
As we also decompose the categorical variables site type, slash-\&-burn intensity and grazed to construct the two design matrices, it is possible to assess their scale-dependent effect on the respective details.
Their smoothed-effect behavior becomes similar to a continuous variable and shows the impact of the increasingly ordered factors.
The coefficients and covariance-function parameters are estimated separately in a joint ML approach for each detail.
Because the number of data points is not excessively large, we can compute the Hessian matrix while optimizing the ML and construct Wald confidence intervals for each estimate.
These Wald confidence intervals are symmetric and therefore they very conservatively imply the significance of the individual estimates.

\subsection{Results}
\begin{figure}[ht]
  \centerline{\includegraphics[width=\textwidth]{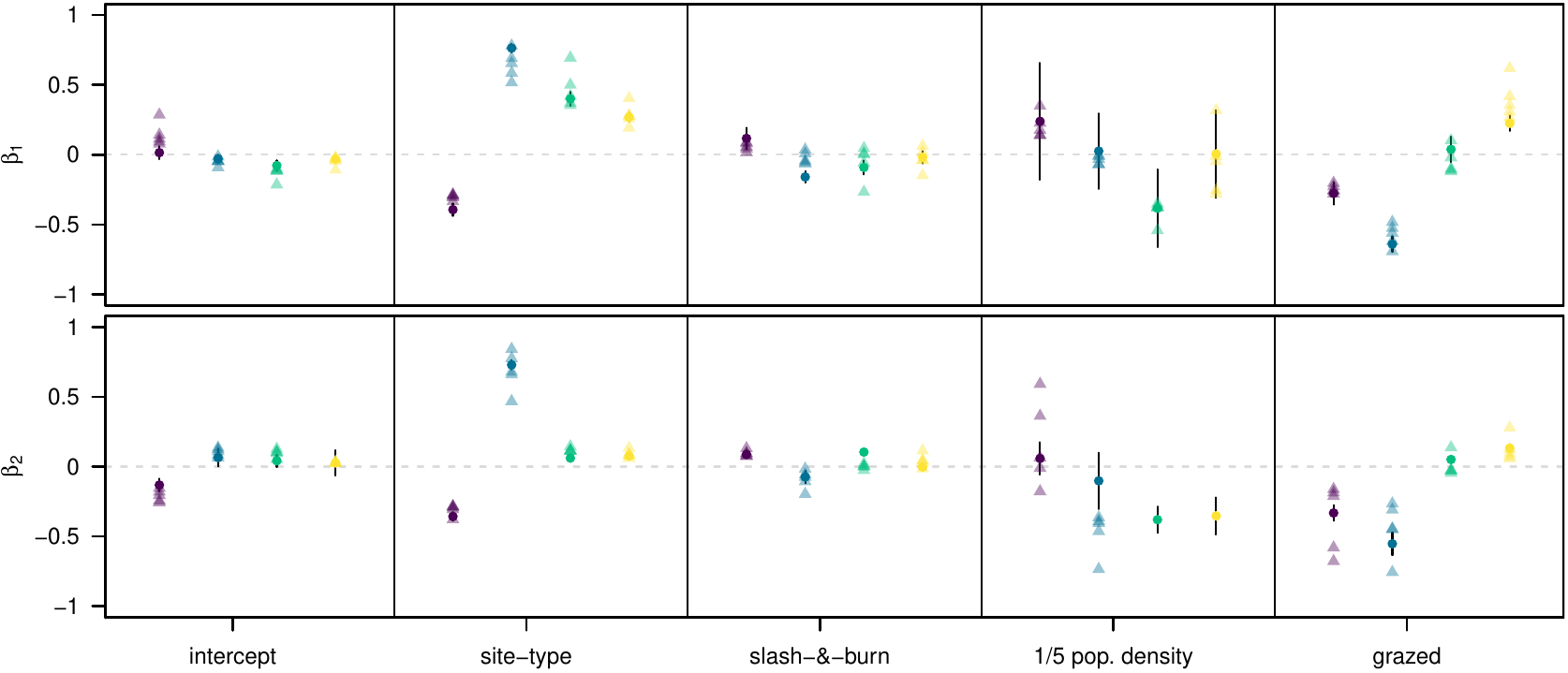}}
  \caption{\small ML estimates of the linear effects of BA detail models: the upper panel is for the $\boldsymbol{z}_1$- and the lower panel for the $\boldsymbol{z}_2$-models.
  In each sub-panel, the colors from left to right are differentiating between pine, spruce, birch and other trees.
  Estimates based on the entire data are visualized by dots, cross-validation estimates with triangles, and 95-\% Wald confidence intervals with black lines.
  The respective estimates of population density are scaled by 1/5 for visualization.}\label{fig:results}
\end{figure}
\begin{figure}[ht]
  \centerline{\includegraphics[width=\textwidth]{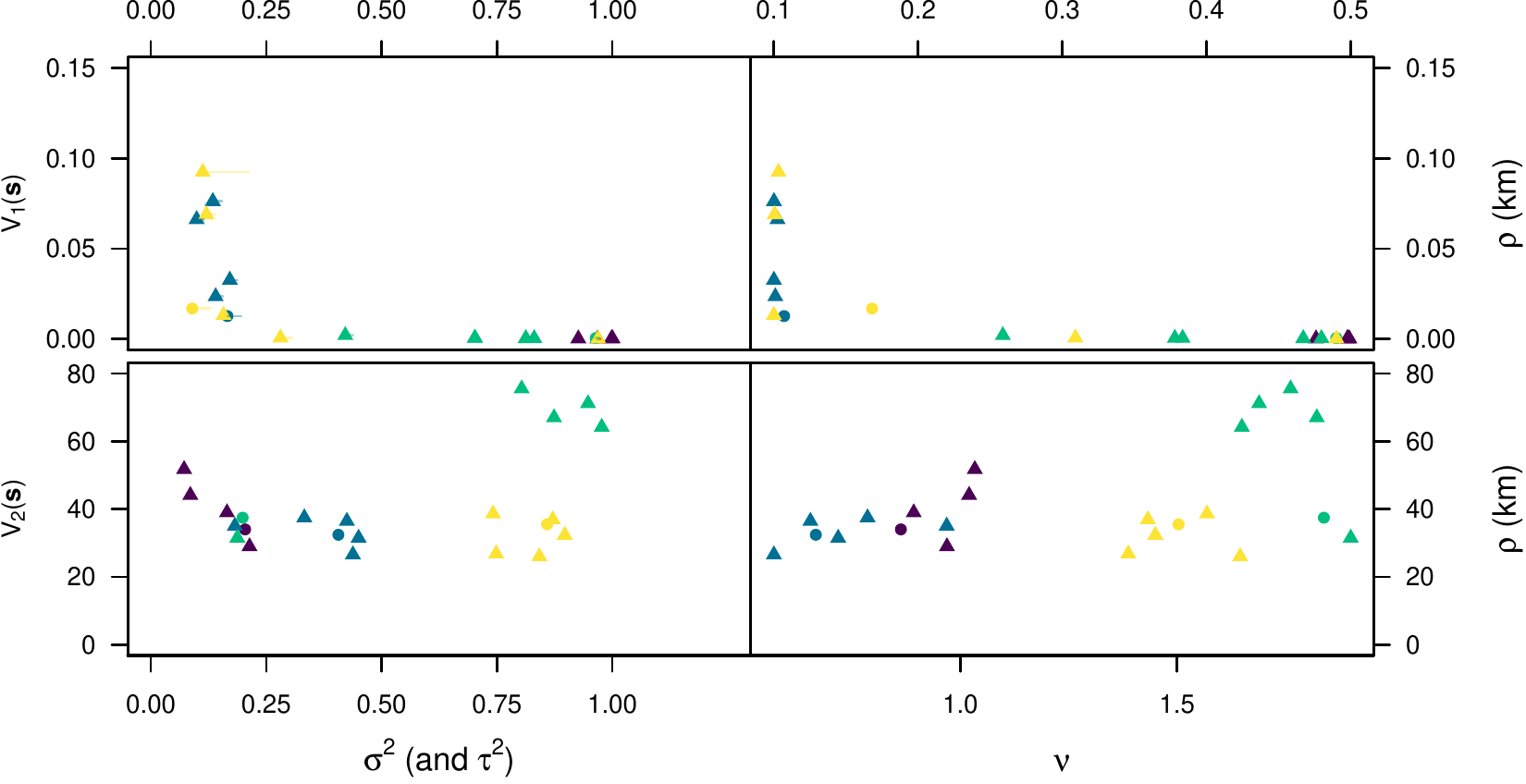}}
  \caption{\small ML estimates of the spatial effects of BA detail models: the upper panel is for the $\boldsymbol{z}_1$- and the lower panel for the $\boldsymbol{z}_2$-models.
  In each sub-panel, the colors from dark to bright are differentiating between pine, spruce, birch and other trees.
  Estimates based on the entire data are visualized by dots and cross-validation estimates with triangles.}\label{fig:spatialresults}
\end{figure}

In the following paragraphs, we describe and interpret the results of the dominant-feature identification, including detail and PW maps as well as the estimates of the linear and the spatial effects.
The resulting scale-dependent linear effects according to Model~\eqref{eq:linearspatialmodel} for pine/spruce/birch/other $\boldsymbol{z}_1$ and $\boldsymbol{z}_2$, including cross-validation estimates, are summarized in Figure~\ref{fig:results}.
The estimates of the spatial components of these models are separately visualized in Figure~\ref{fig:spatialresults}.

Comparing overall the details $\boldsymbol{z}_1$ and $\boldsymbol{z}_2$ with the scale-space decomposition Figure~\ref{fig:fl_pine} and in the Supplementary Material Figures~\ref{fig:fl_spruce}, \ref{fig:fl_birch} and \ref{fig:fl_other}, we can already observe that the small-scale details ($\boldsymbol{z}_1$'s) mainly contain plot-to-plot variation.
The large-scale details show for all species regional scale-dependent dominant features, and it seems that spruce BA has larger features than any of the other tree species.

We assessed scale-dependent linear effects of the four variables visualized in Figure~\ref{fig:fl_data2} on the scale-dependent BA details.
The intercept, the mean value of the decomposed BAs is for all tree species close to zero, and the respective small- and large-scale details of all tree species nearly cancel each other out.

The site-type effects show the expected ecological demeanor, consistently on $\boldsymbol{z}_1$ and $\boldsymbol{z}_2$.
Pine BAs are negatively associated with increasing richer site types, which is consistent with pine being more abundant on poor sites.
Spruce, birch and other deciduous trees that typically dominate more fertile sites show the opposite relationship to pine; i.e., the richer the site type, the higher the BAs.
These effects are consistent for coniferous trees on small and large scales, but for deciduous trees much weaker on $\boldsymbol{z}_2$.

Pine BAs are on $\boldsymbol{z}_1$ and $\boldsymbol{z}_2$ consistently positively associated with increasing intensity in slash-\&-burn agriculture.
By contrast, spruce BAs are consistently negatively related on both small and large scales with this variable, in alignment with the hypothesis that this practice was detrimental for spruce.
Deciduous trees BAs are negatively influenced on $\boldsymbol{z_1}$ from this form of cultivation.
However, on $\boldsymbol{z}_2$, the effect on birch and other deciduous trees BAs becomes positive.
The stronger positive association of birch BAs can be explained by the fact that mountain birches are highly abundant in the north of Finland and were not exposed to slash-\&-burn agriculture in this region.
Moreover, birch usually reappears after a quicker recovery phase as spruce, which explains the different effect on $\boldsymbol{z}_2$ compared to spruce.
On $\boldsymbol{z_1}$ of other deciduous trees, BAs are slightly negatively associated with this practice.
However, on the large scale this association becomes positive, as hypothesized.

Pine BAs are consistently positively associated with population density.
However, this effect is diminishing on the large scale.
Spruce and other deciduous trees are showing a slightly positive effect on $\boldsymbol{z}_1$, but a much stronger negative effect on $\boldsymbol{z}_2$.
Birch BAs are consistently negatively associated with population density, and other deciduous trees show no effect on $\boldsymbol{z}_1$ but a strongly negative association on $\boldsymbol{z}_2$.

Moreover, on $\boldsymbol{z}_1$ and $\boldsymbol{z}_2$ BAs of both conifer tree species are negatively affected by grazing.
Especially spruce could severely suffer from grazing as its strategy in natural conditions is to regenerate under a deciduous canopy and gradually take over site dominance.
Seedlings also suffer from cattle trampling.
Birch BAs are consistently marginally positively associated.
In comparison, BAs of other deciduous trees show a more substantial positive effect.  Deciduous trees have a pioneer strategy, i.e., they regenerate quickly on disturbed sites.
However, cattle were often let into forest stands for grazing, which was in the recovery phase of slash-\&-burn cultivation, and their grazing preferences further shaped the species composition.
In particular, this led to the increase in alder that is poorly palatable for the cattel.

Comparing the top and bottom panels in Figure~\ref{fig:spatialresults}, it is evident that the respective $\boldsymbol{z}_1$ shows plot-to-plot variation in BA for each tree species studied.
The respective estimated effective ranges on $\boldsymbol{z}_1$ are all smaller than the constructed distances between sample plots of 2~km.
The estimated nugget effects ($\tau^2$) are substantial for most of the estimated partial sill, so there is still considerable location-independent variation in these details.
The smoothing parameters are either estimated to be less than~$0.5$, or the effective range is effectively zero.
These results consistently imply for all tree species that there is little to no spatial correlation on $\boldsymbol{z}_1$.
We detect scale-dependent dominant features on $\boldsymbol{z}_2$, which have effective ranges of approximately 40~km.
These show no nugget effect on $\boldsymbol{z}_2$, and the smoothing estimates imply that there are connected patterns on these details.
This means that we can identify variation at a regional scale.

When we compare the spatial effects between the different tree species in Figure~\ref{fig:spatialresults}, it is apparent that the estimates form four almost disjoint clusters, reflecting the differences between the respective spatial-dependency characteristics.
Thereby the effective ranges are similar, but the variance and smoothness parameter differ.
This is well reflected in the respective scale-dependent features shown in Figure~\ref{fig:fl_pine} and Figures~\ref{fig:fl_spruce}, \ref{fig:fl_birch} and \ref{fig:fl_other} in the Supplementary Material, with equally sized large features.
The features of spruce and pine BAs are, however, less smooth then the others features.

Overall, cross-validation estimates in Figure~\ref{fig:results} show that the linear effects are very stable.
The cross-validation estimates of the spatial effects in Figure~\ref{fig:spatialresults} show that at most some regional artifacts affect the partial sill.
However, the estimated effective ranges and smoothness estimates are evidently very stable.
The assumption of isotropic spatial processes is therefore supported, and overall we can not detect any sign of overfitting.
We calculated also the RMSE and CRPS for each training set, predicting to $\boldsymbol{x}$ from the sum of training $\boldsymbol{z}_1$ and $\boldsymbol{z}_2$.
Table~\ref{tab:cverrors} of the Supplementary Material shows the respective values.
However, as this complete cross-validation includes also the scale selection, the predictions are additionally biased and so are their quality measures.

\section{Discussion}\label{sec:discussion}
Extending the feature identification for geostatistical data fills a major gap in spatial statistics and enables its application to a vast area of data.
The outlined method relies on the ML estimates of the Mat\'ern covariance parameter, limiting the method to moderate data sizes.
However, many approximations are provided in spatial statistics, which can be plugged in at the necessary steps to overcome this limitation.
Extending the feature attributes of scale-dependent features to additional linear effects offers further insight into possible drivers of the underlying processes.
We demonstrated that using scale-dependent predictor variables, different effects on different scales might be found. Especially predictors with large scales are likely to contain hidden small-scale effects.

Applying the dominant-feature identification to the NFI1 BAs, we showed that there are two essential scales with interesting differences, the small plot-to-plot variation scale and the large regional scale.
The results imply that variation in BAs is driven by differences in site-type and anthropogenic variables on both scales.
In general, the effect of site type is considerably stronger than slash-\&-burn, which influenced large areas of the Finnish landscapes in the past.
This might be also because this method was applied predominantly on richer soils.
The high plot-to-plot variation in the site type is typical for Fennoscandian boreal landscapes, which form a matrix of forests on varying edaphic conditions (i.e., bedrock, quaternary deposits) on mineral soils, open and forested peatlands, and lakes and other water bodies.
Known phenomena such as the decline of spruce and the increase of other deciduous trees associated with slash-\&-burn were confirmed. 
We also found an increase in grazing areas common for other deciduous trees on the large scale.
As cattle were often let into areas after slash-\&-burn cultivation was practiced, there is a possible interaction effect between these two variables.
The strongest effect has population density, and we showed the expected decrease in pine BA at the regional scale.
However, it is not entirely clear how the interpolation/smoothing used while digitizing the variable affects this result.
In conclusion, with these scale-dependent models, it is possible to capture differentially directed associations between predictors and BAs, for example, between slash-\&-burn and birch, which show different behavior at small and large scales.
This would be impossible to assess with conventional models.

\section{Contributions}
RF: Conceptualization, Methodology, Application, Data curation, Visualization, Software, Writing - original draft, review \& editing.
TA: Data source \& curation, Application, Ecologial background, Writing – review \& editing.
LR: Methodology, Application, Writing – review \& editing.
TK: Data source, Writing – review \& editing.
RF: Methodology, Supervision, Funding acquisition, Writing - review \& editing.

\section{Acknowledgments}
The authors thank Agata Guirard, Lucas Kook and Michael Hediger for the stimulating discussions during the development of this work.
We also thank the IT team of the Department of Mathematics of the University of Zurich for their excellent support.
This work is supported by a GRC Travel Grant from the University of Zurich and the Swiss National Science Foundation through grant SNSF-175529.

{\renewcommand\baselinestretch{1.05}
\small
\bibliographystyle{mywiley}
\bibliography{processdecomp.bib}}

\appendix

\section{Source Files}\label{appen:code}
  Supplementary material is available in the git repository \url{https://git.math.uzh.ch/roflur/dominantfeaturesinfinnishforestdata}.
  It contains the following files:
\begin{itemize}
  \item README.md: detailed description of available \textsc{R}-code.
  \item LICENSE: GNU general public license.
  \item source/: contains the \textsc{R}-devel package \textbf{mresa} to run the analysis.
  \item analysis/: contains \textsc{R}-scrips to run feature identification of Finnish forest inventory data.
\end{itemize}

\section{Smoothing-correlation parameter}\label{sec:scaleselection}
We use simulated data to illustrate how the tuning smoothing-correlation parameters in $\boldsymbol{\widetilde{S}}_{\lambda}$ influence the decomposition and corresponding detail processes.
We construct a composition of two fields by sampling from two zero-mean Gaussian processes, using different range and smoothing parameters for a Mat\'ern covariance function ($\rho_1 = 0.05,\, \nu_1 = 0.8$ and $\rho_2 = 0.2,\, \nu_2 = 2.2$).
The sampling domain for these processes is the unit square $[0,1] \times [0,1]$ with $2^{10}$ uniform randomly sampled locations.

We calculate the scale derivative for a sequence of range and smoothness parameters, choose the scale according to the local minima with respect to the norm of the maximum for each scale derivative and estimate the detail process parameters.
\begin{figure}[ht]
  \centerline{\includegraphics[width=\textwidth]{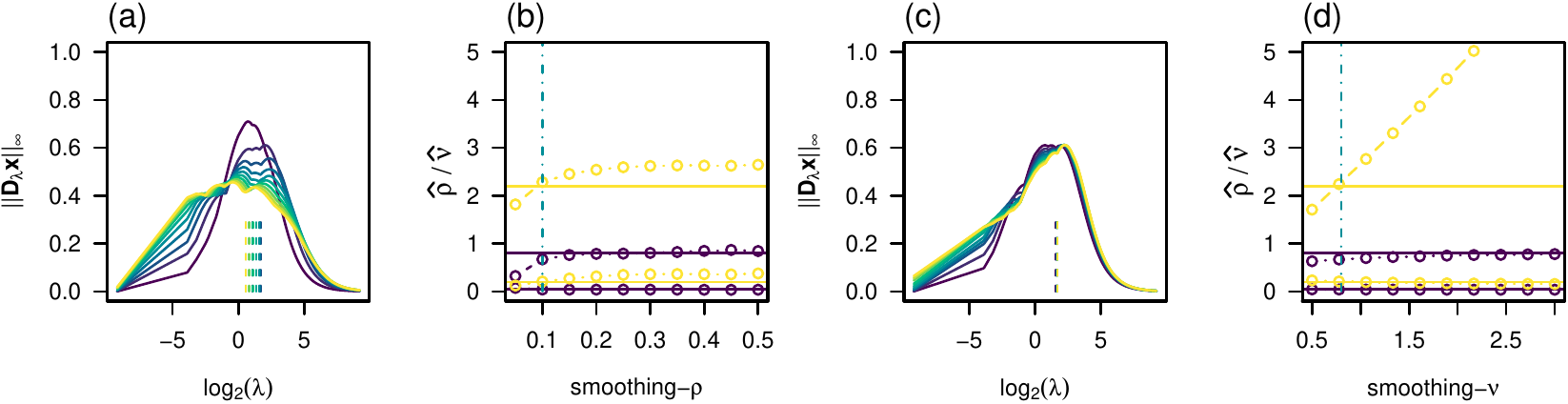}}
  \caption{\small Smoothing-correlation simulations: (a) and (c) scale derivatives for different ranges and smoothness parameters --- the brighter the color of the scale derivative the higher the parameter;
  (b) and (d) effective range and smoothness ML estimates of detail process $\boldsymbol{z}_1$ (dark color) and $\boldsymbol{z}_2$ (bright color) for different ranges/smoothness --- solid horizontal lines depict the true parameters and dashed vertical lines show the optimal choices.}
  \label{fig:test_scales_maternC}
\end{figure}

Panel (a) and (c) of Figure~\ref{fig:test_scales_maternC} show the demeanor of the scale derivative.
It is apparent that with increasing range and increasing smoothness parameter the minima become stable at some point.
Panel (b) and (d) of this figure visualize the same behavior for the parameter estimates of the detail processes; at some point, they stop changin, except for the smoothness of $\boldsymbol{z}_2$ process, which seems to increase with increasing smoothness parameter.
Therefore, if the estimated smoothness parameter of detail $\boldsymbol{z}_{L-1}$ grows unreasonably large, this indicates that the smoothness parameter of the smoothing-correlation function is chosen too large.
These results show that the choice of the range and smoothness do not hamper the choice of the scale or of the detail process.
However, for computational reasons, we propose to choose these parameters to be as small as possible.

\section{Supplementary to Section ``Identifying dominant features in Finnish forest inventory records''}\label{app:flplots}

\begin{figure}[htb]
  \centerline{\includegraphics[width=\textwidth]{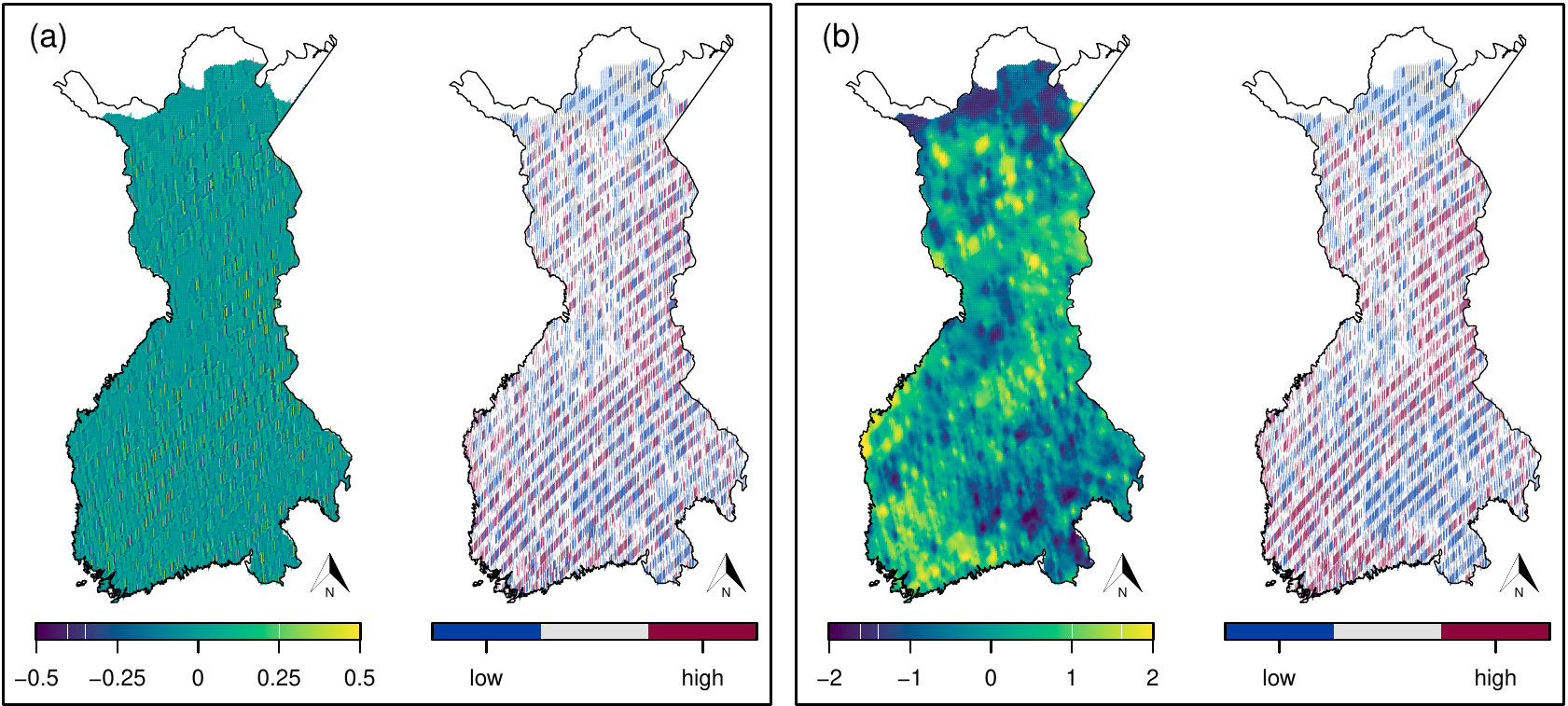}}
  \caption{\small Scale-space decomposition of pine BA: (a) spruce $\boldsymbol{z}_1$ and spruce PW$_1$; (b) spruce $\boldsymbol{z}_2$ and spruce PW$_2$.}
  \label{fig:fl_spruce}
\end{figure}

\begin{figure}[htb]
  \centerline{\includegraphics[width=\textwidth]{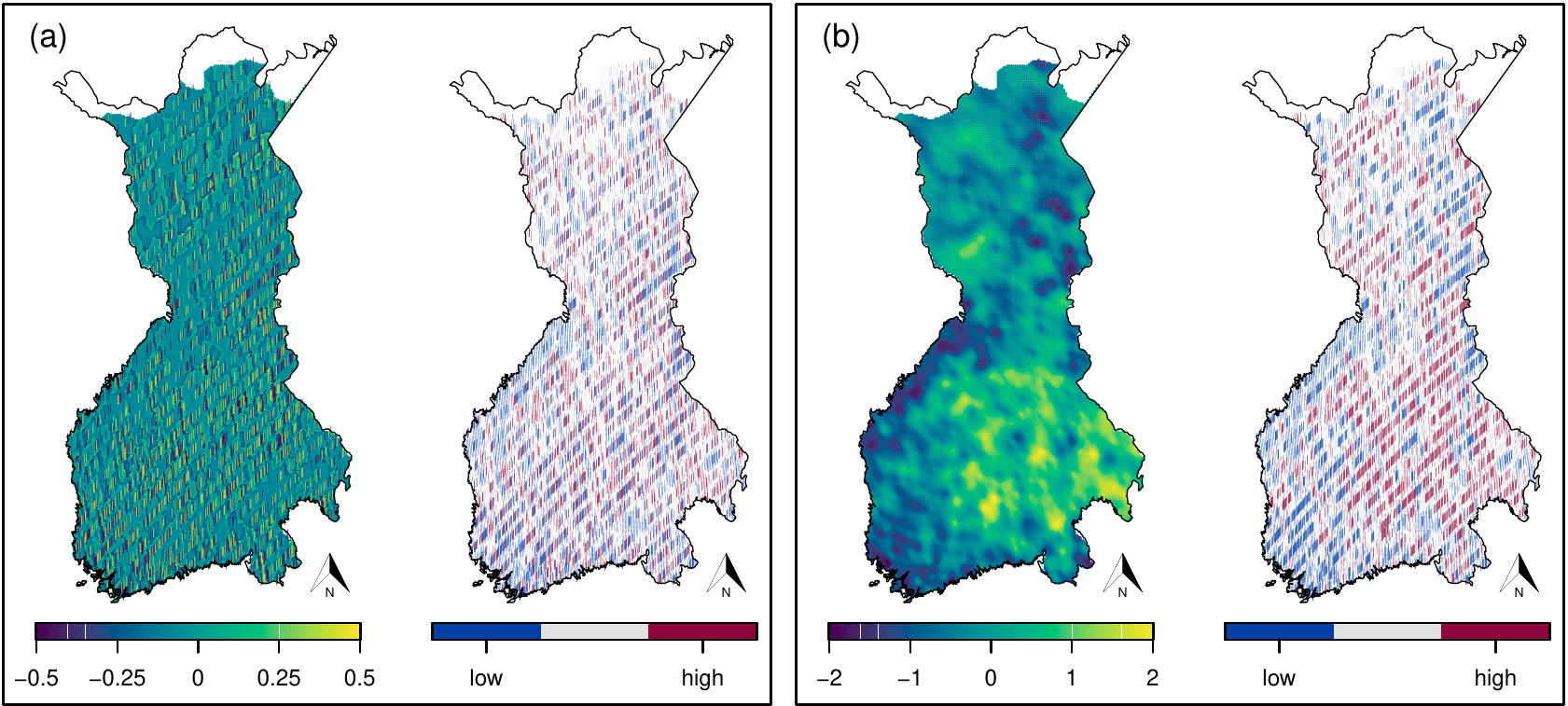}}
  \caption{\small Scale-space decomposition of birch BA: (a) birch $\boldsymbol{z}_1$ and birch PW$_1$; (b) birch $\boldsymbol{z}_2$ and birch PW$_2$.}
  \label{fig:fl_birch}
\end{figure}

\begin{figure}[htb]
  \centerline{\includegraphics[width=\textwidth]{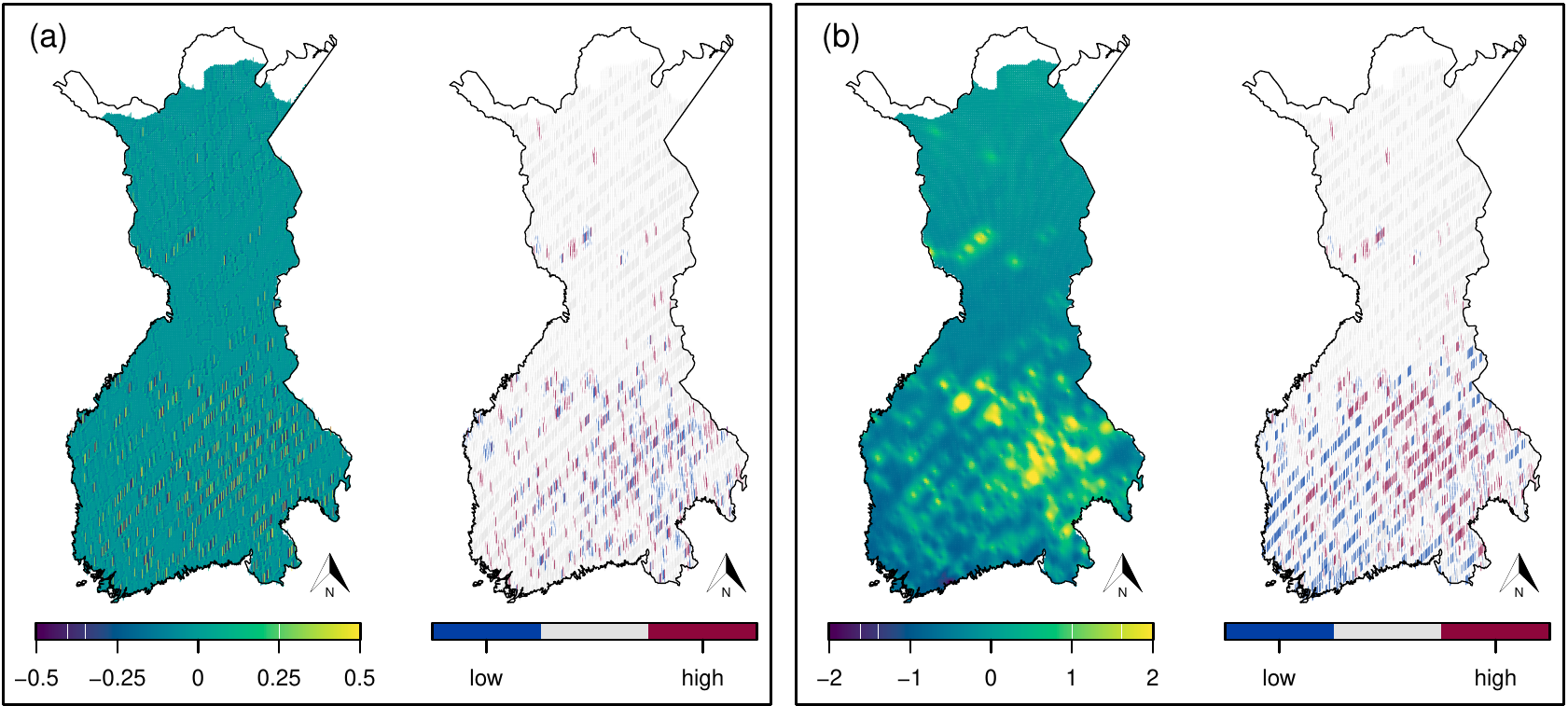}}
  \caption{\small Scale-space decomposition of other BA: (a) other $\boldsymbol{z}_1$ and other PW$_1$; (b) other $\boldsymbol{z}_2$ and other PW$_2$.}
  \label{fig:fl_other}
\end{figure}

\begin{center}
\begin{table}
\caption{Cross-validation prediction errors: RMSE and CRPS from predictions based on different training sets for each tree species.}
\label{tab:cverrors}
\begin{tabular}{c|rrrr|rrrr}
   & \multicolumn{4}{c|}{ RMSE } & \multicolumn{4}{c}{ CRPS }  \\
  set/species & pine & spruce & birch & other & pine & spruce & birch & other \\ \hline
  1 & 0.913 & 0.970 & 1.075 & 0.912 & 0.852 & 0.606 & 0.634 & 0.703 \\ 
  2 & 0.814 & 0.970 & 1.035 & 0.984 & 0.821 & 0.599 & 0.670 & 0.870 \\ 
  3 & 0.819 & 0.989 & 1.013 & 0.880 & 0.699 & 0.662 & 0.665 & 0.763 \\ 
  4 & 0.892 & 1.002 & 1.246 & 0.937 & 0.860 & 0.655 & 0.578 & 0.820 \\ 
  5 & 0.792 & 0.928 & 1.075 & 0.870 & 0.836 & 0.619 & 0.703 & 0.809 \\ 
  \end{tabular}
\end{table}
\end{center}

\end{document}